\documentclass[pra,twocolumn,floatfix,superscriptaddress,showpacs]{revtex4}
\usepackage{graphicx}
\usepackage{amsfonts}
\usepackage{amssymb}
\usepackage{amsmath}
\usepackage{mathrsfs}
\usepackage{color}
\usepackage[normalem]{ulem}

\makeatletter

\newcommand{\Rmnum}[1]{\expandafter\@slowromancap\romannumeral #1@}
\makeatother
\begin{document}
\title{Dissipative Transport of Trapped Bose-Einstein Condensates
  through Disorder} 
\author{S.~G.~Bhongale}
\affiliation{Department of Physics and Astronomy \& Rice Quantum
  Institute, Rice University, Houston, TX 77005, USA}
\affiliation{Department of Physics \& Astronomy, George Mason University, , MS 3F3, Fairfax, VA 22030, USA}
\author{Paata Kakashvili} 
\affiliation{Department of Physics and Astronomy \& Rice Quantum
  Institute, Rice University, Houston, TX 77005, USA}
\affiliation{NORDITA, Roslagstullsbacken 23, 106 91 Stockholm, Sweden} 
\author{C.~J.~Bolech}
\affiliation{Department of Physics and Astronomy \& Rice Quantum
  Institute, Rice University, Houston, TX 77005, USA}
\affiliation{Department of Physics, University of
  Cincinnati, Cincinnati, OH 45221, USA}
\author{H.~Pu} 
\affiliation{Department of Physics and Astronomy \& Rice Quantum
  Institute, Rice University, Houston, TX 77005, USA}

%\maketitle
%\begin{article}
\begin{abstract}
After almost half a century since the work of Anderson [Phys. Rev. {\bf 109}, 1492 (1958)], at present there is no well established
theoretical framework for understanding the dynamics of interacting particles in the presence of disorder.
Here, we address this problem for interacting bosons near $T=0$, a situation that has been realized
in trapped atomic experiments with an optical speckle disorder.
We develop a theoretical model for understanding the hydrodynamic
transport of \emph{finite-size} Bose-Einstein condensates through
disorder potentials. The goal has been to set up a simple model that will retain
all the richness of the system, yet provide analytic expressions,
allowing deeper insight into the physical mechanism. Comparison of our theoretical predictions with
the experimental data on large-amplitude dipole oscillations of a
condensate in an optical-speckle disorder shows striking agreement. We are able to quantify various dissipative
regimes of slow and fast damping. Our calculations provide a clear evidence of reduction in disorder strength due
to interactions. The analytic treatment presented here allows us to predict the power law governing the interaction
dependance of damping. The corresponding exponents are found to depend sensitively on the dimensionality and are in excellent agreement with experimental observations.
Thus, the adeptness of our model, to correctly
capture the essential physics of dissipation in such transport
experiments, is established.
\end{abstract}
\date{today}
\maketitle
\section{Introduction}
Bose-Einstein condensates generated in trapped
atom experiments constitute mesoscopic quantum objects that are well
localized in space and exhibit quantum properties such as interference
and phase coherence. They represent a highly controllable and
potentially rich template for addressing fundamental questions
related to quantum measurement and configuring precision measurement
devices \cite{bhongale}. Here we study the unique problem of transporting such
finite-sized BEC's on rough surfaces, a situation inevitably encountered with
BEC's generated on chips. There the roughness originates from the
fluctuations of the surface fields \cite{kruger}.

In a broader context, transport of superfluids through
disordered potentials has always been a topic of theoretical intrigue
ever since the first experimental observations concerning super-flow
of Helium-4 through porous media \cite{superflow}. Huang and Meng
first proposed a hard-sphere Bose gas in a random potential as a model
describing the static properties of this scenario \cite{huang}. Even
though mostly qualitative, this work was very insightful and provided
definite clues to the connection between boson localization and
condensate depletion.  Subsequently, using a quantum hydrodynamic approach, Giorgini \emph{et al.} \cite{giorgini}
looked at the disorder induced phonon damping in a Bose superfluid. However,
not much else was said about dynamics. In fact, after almost two decades, questions
related to the transport of superfluids through disorder have hardly
been touched mostly due to the lack of proper experimental
setups. However, very recent, highly precise experiments on trapped
atoms \cite{randy,randy2,demarco} have begun to expose puzzling new
aspects of such transport through disorder. A sharp reduction in
dissipation rate beyond the Landau critical velocity \cite{landau} is
observed. Further, the experiments also indicate a power-law
dependance of the dissipation rate on the interaction strength. Finally, when we view these
disordered BEC experiments as well localized
macroscopic quantum objects sliding on rough surfaces, it touches
upon another interesting research area with open questions related to
the origins of non-linear friction laws \cite{friction}.

Trapped atomic systems constitute a promising new platform whereby the
many-body physics of various disorder-induced phenomena may be
understood. Contrary to condensed matter, these atomic
systems are intrinsically pure and allow for controlled insertion of
disorder by, {\it e.g.}, modulating the lattice potential
\cite{massimo}, or creating an optical speckle pattern
\cite{randy,aspect}. Experimentalists are even able to control, to a
good extent, the correlation properties of these disorder
potentials.

The significance of this type of extreme control needs no further
discussion if we simply note that, for the first time, experiments are
able to provide explicit access to the localized wavefunction, thus
allowing direct evidence of Anderson localization, an exquisite
phenomenon responsible for the complete disappearance of metallic
conductivity \cite{massimo,aspect,anderson}.  Such spectacular
developments, with no surprise, give compelling reasons to
theoretically investigate the disordered Bose system in the light of
the new techniques of probing made available by ultra-cold
atoms. Especially, with the possibility of tuning the inter-particle
interactions via techniques of Feshbach resonance \cite{feshbach}, ultra-cold atom
experiments provide a systematic way to investigate the role of
interactions in the localization phenomenon; an aspect that has
generated much debate in the field. Numerous transport experiments can
be performed, requiring new insights due to the finite spatial extent
of these trapped-atom condensates.
In addition, excitations in a finite BEC are fundamentally different
in symmetry from those in the translationally invariant counterpart.
Unfortunately, at present there is no well established
theoretical framework for describing such non-equilibrium hydrodynamic
scenarios. Experimental observations \cite{randy,randy2,demarco}
continue to demand a clear understanding of the physical mechanism.
Recent numerical simulations based on the Gross-Pitaevskii equation
\cite{albert}, confirm some of these observations and provide some
hints towards the possible physical mechanism, but a more detailed
theoretical model is missing. As will be shown below, our model provides a
microscopic underpinning of the experimental and numerical
results. This allows us to determine exponents governing the power law
dependence of the dissipation on the scattering length, giving a
deeper understanding of the interplay between disorder and
interactions. Also, as will be evident from our discussion, such exponents
critically depend on dimensionality.

\begin{figure}[h]
\includegraphics[scale=.37]{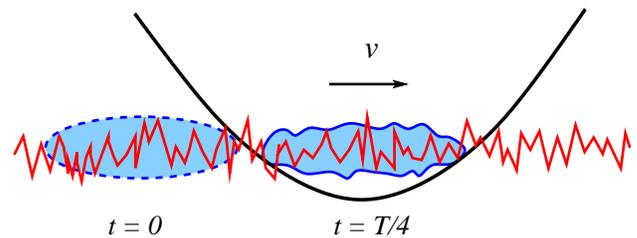}
\caption{(Color online) Sketch of the experimental scenario in question. The zigzag
    line corresponds to the speckle potential that is superimposed
    onto the smooth harmonic potential. The condensate (shown in blue)
    is shifted to one edge of the trap and released at $t=0$,
    resulting in dipole oscillations.}
\label{schematic}
\end{figure}

In this paper, we present a theory for determining the transport
properties of a trapped BEC flowing through a disordered potential,
similar to that used in the experiments of
Refs.~\cite{randy,randy2,aspect}. In the past, similar interest in
transport of localized wavepackets, for example in microwaves
\cite{tiggelen}, has been limited to time scales of the order of
the diffusion time $\tau$. While our proposed framework may also be applied
to this regime as well, we intend to focus on the specific
experimental situation discussed in Refs.~\cite{randy,randy2}.
Here, the dissipative transport is achieved by generating a center-of-mass
mode of the trapped condensate and allowing it to propagate through an
optical speckle potential (see Fig.~\ref{schematic}). The damping of this mode
 is measured as a function of interaction strength and the
center-of-mass velocity.
The strength of the disorder is such that the period of oscillations $T \ll \tau$.

\section{Collective Modes (Excitations) of a Trapped BEC}
A tightly confined BEC can be made to oscillate as a whole, for
instance, by abruptly shifting the trap by a small distance.
If the displacement is small, these excitations can be easily
described by linearizing the Gross-Pitaevskii (GP) equation
\begin{eqnarray}
  i\hbar\frac{\partial}{\partial t} \psi_0({\bf r},t)=\left[-\frac{\hbar^2\nabla^2}{2m}+V_{{\rm ext}}({\bf r})+\lambda |\psi_0({\bf r},t)|^2\right]\psi_0({\bf r},t),&&\nonumber
\end{eqnarray}
where $V_{{\rm ext}}({\bf r})$ is the trapping potential, and the
interaction parameter $\lambda$, in terms of the $s$-wave atom-atom
scattering length, $a$, is $\lambda=4\pi\hbar^2a/m$. The set of
Bogoliubov equations that follow provide the eigenenergies,
$\Lambda_i$, of the elementary excitations. As is apparent from the above GP equation, interactions play an important role in defining these excitations.

Among other types of collective oscillations exhibited by the trapped
condensates, of special significance is the dipole mode.  It
corresponds to the center of mass motion of the condensate which due
to the harmonic confinement has the same frequency as the trap and is
unaltered by interactions. In general, for 3D confinement, there exists
a whole set of low energy modes that can be obtained analytically by
solving the coupled hydrodynamic equations (derived from the GP
equation above) for the velocity, ${\bf v}({\bf r},t)$, and the
density, $n({\bf r},t)$:
%\begin{eqnarray}
$\partial n/\partial
    t+{\rm div}({\bf v}n)=0$ and
$m\partial {\bf v}/\partial t +\nabla
  \left\{m{\bf v}^2/2+V_{{\rm ext}}+\lambda n\right\}=0$.
%\end{eqnarray}
These equations allow for
solutions, where the condensate is oscillating as a whole
\cite{stringari}, with frequencies given by
%\begin{eqnarray}
$\omega_e=\omega_{{\rm ho}}\sqrt{2n_r^2+2n_r\ell+3n_r+\ell}$,
%\end{eqnarray}
assuming $V_{{\rm ext}}({\bf r})$ is harmonic with frequency
$\omega_{{\rm ho}}$. The dipole mode is the one with quantum numbers
$(n_r=0,\ell=1)$.  In the absence of any other potential (disordered
or not), the harmonic trap is special, leaving the center-of-mass
motion completely decoupled from the relative degrees of freedom even
for large amplitudes. However this feature is immediately lost as soon
as any other external potential, {\it e.g.} disorder, is turned on,
resulting in coupling of different modes, eventually leading to the
complete decay of a well defined oscillating mode.

\section{Key Concepts} Our model consists of two main ingredients, \emph{disorder}
and \emph{inhomogeneity}, which we describe below. \\
\subsection{Disorder} For treating disorder within an analytic
approach, among several well established techniques in the literature
\cite{replica,methods}, we find it convenient to use the \emph{replica} method.
The disorder potential $U_{{\rm d}}({\bf x})$ is described
in terms of a Gaussian distribution function with a strength $\gamma$
such that
%\begin{eqnarray}
$P[U_{{\rm d}}]=\exp [-1/(2\gamma^2)\int {\rm d}{\bf
  x}{\rm d}{\bf x}'U_{{\rm d}}({\bf x})K^{-1}({\bf x}-{\bf
  x}') U_{{\rm d}}({\bf x}')]$
%\end{eqnarray}
 and $K$ describes the spatial
correlations \cite{replica}. Such procedure automatically allows for a
freedom in choosing the actual disorder potential, the details of
which are quickly erased by multiple scattering. To further simplify
our investigation, we consider a white noise correlation function
$K({\bf x})\sim \delta({\bf x})$. Thus,
%\begin{eqnarray}
$\langle U_{{\rm d}}({\bf
  x})\rangle_{{\rm dis}}=0$ and $\langle U_{{\rm d}}({\bf
  x})U_{{\rm d}}({\bf x}')\rangle_{{\rm dis}}=\gamma^2\delta({\bf
  x}-{\bf x}')$,
%\end{eqnarray}
where $\langle\cdot\cdot\cdot\rangle_{{\rm dis}}$ means averaging
over disorder. However, the \emph{trick} of using the \emph{replica}
method lies in following a particular procedure for taking the
sequence of averages (quantum followed by disorder), essential for
calculating observables.
While the \emph{replica trick} has been typically used in relation to
Anderson localization of non-interacting electrons,
%this will be the first instance where
we adopt it in the context of ultra-cold
interacting atoms. A brief outline of this technique is provided in
the Appendix. Aside from technical aspects,
the \emph{replica trick} allows for a systematic diagrammatic
perturbation expansion within which all transport
properties can be calculated, at least in principle.\\

\subsection{Inhomogeneity} Next we use local density
approximation (LDA) \cite{stringari} to account for the inhomogeneous density
distribution of the trapped condensate. This is achieved by defining a position
dependent chemical potential,
\begin{eqnarray}
\mu\rightarrow \mu({\bf
  r})&=&\mu-V_{{\rm trap}}({\bf r}),\label{lda1}
\end{eqnarray}
clearly justified in the Thomas-Fermi limit \cite{stringari}, which is typically always the case
for trapped-atom BEC experiments. Thus, the density of the stationary
condensate in the absence of disorder is given by
\begin{eqnarray}
\rho({\bf r})=\mu({\bf r})/\lambda\label{lda2}
\end{eqnarray}
Further, we assume that as the condensate undergoes dipole
oscillations, on average the shape of the condensate is retained,
meaning $\rho({\bf r}+{\bf r_{\rm cm}}(t))=\rho({\bf r})$. This is
an extremely good assumption, it stems from the rigidity associated with
the Bose condensate.
% The LDA has been proved to work prefectly in
% almost all trapped atomic experiments.
This greatly simplifies the
problem by allowing us to define all the relevant quantities locally,
for example we define the local Bogoliubov excitation energies \cite{ripka}
\begin{eqnarray}
\Lambda_{{\bf k}}[{\bf r}]=\sqrt{e_k^2-\lambda^2\rho({\bf r})^2}\label{bdglocal}
\end{eqnarray}
with $e_k=\hbar^2k^2/2m+\lambda\rho({\bf r})$. Also, from the linear
part of the above dispersion relation, we can define a local sound
velocity\\
\begin{eqnarray}
c[{\bf r}]=\sqrt{\lambda\rho({\bf r})/m}.\nonumber
\end{eqnarray}
\begin{figure}[ht]
  \includegraphics[scale=.42]{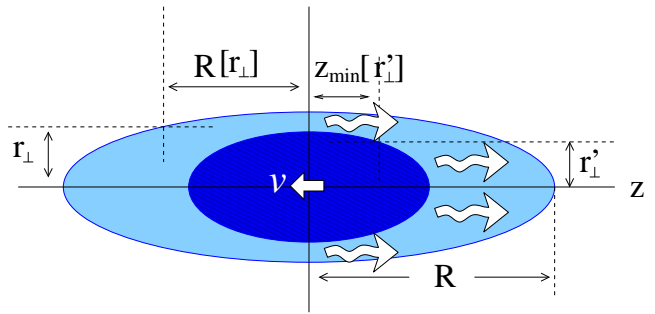}
  \caption{(Color online) Sketch showing the different relevant distances defined
    in the text. Dark (light) blue color represent spatial regions of
    the condensate where the local sound velocity is greater (lesser)
    than the center-of-mass velocity. Graphic arrows indicate excitation moving
    along the $z$ direction.}
\label{distances}
\end{figure}

\section{The Model} The total Hamiltonian for the problem can be
written as the sum
\begin{eqnarray}
\hat{H}&=&\underbrace{\hat{H}_{{\rm free}}+\hat{H}_{{\rm int}}}+\hat{V}_{{\rm trap}}+\hat{U}_{{\rm d}}\nonumber\\
&=&\hspace{0.74cm}\underbrace{\hat{H}_{{\rm eff}}\hspace{0.45cm}+\hat{V}_{{\rm trap}}}+\underbrace{\hat{U}_{{\rm d}}}\nonumber\\
&&\hspace{1.4cm}{\rm LDA}\hspace{1cm}{\rm replica},\nonumber
\end{eqnarray}
where $\hat{H}_{{\rm free}}$ is for the free non-interacting atoms
and $\hat{H}_{{\rm int}}$ is due to the atom-atom interaction. As
indicated above, we first write an effective Hamiltonian for the
bosons in the absence of any potentials by treating
$\hat{H}_{{\rm int}}$ at the mean field level. Diagonalizing
$\hat{H}_{{\rm eff}}$, the lowest eigenvalue corresponds to the
condensate mode at temperatures $T<T_{{\rm BEC}}$. The excitations
above the ground state are nothing but the Bogoliubov modes described
above and are orthogonal to the condensate mode. Any other external
potential will in general couple these different eigenmodes. However,
as mentioned earlier, the trapping potential is typically smooth and
LDA holds, drastically simplifying the calculation. The only piece that
remains to be included is the potential, $\hat{U}_{{\rm d}}$. At any
spatial location, the disorder tends to deplete the condensate by
populating the Bogoliubov modes. However, this does not happen if the
condensate is at rest, with zero center-of-mass velocity. Due to the superfluid
property of the BEC, Landau criterion holds and there is a non-zero
critical velocity below which no excitations can be generated and the
nonequilibrium state will persist forever. This is precisely where
the finite size of the condensate plays an important role. At any
given center-of-mass velocity $v$, on the contrary, there is always a finite
region of the condensate, shown by the light blue color in
Fig.~\ref{distances}, where $v>c[{\bf r}]$ thus satisfying Landau
criterion for excitations and providing a channel for the energy to be
transferred from the center-of-mass motion to the Bogoliubov
modes. In essence, this is the physical mechanism that provides the
frictional force responsible for slowing down the condensate center-of-mass
motion.\\
\textbf{Remark:}
\emph{Before we proceed to calculations, we need to take
into account the fact that, previously, excitations were calculated in
the rest frame of the condensate. Thus, if we wish to continue
calculating everything else in the same frame, it implies having to
use a time dependent disorder potential.}\\
This is certainly a
non-trivial complication and cannot be simply neglected. Fortunately,
LDA turns out to be perfectly suited for such a situation. It
allows us to work in the lab frame instead, hence static disorder. Now, from Eq.~\eqref{bdglocal}, the locally defined
Bogoliubov modes are eigenstates of momentum and hence remain
unaltered under a Galilean frame transformation, only their energies are
shifted according to:
\begin{eqnarray}
\Lambda_{\bf
  k}\rightarrow\Lambda_{\bf k}^{{\rm lab}}&=&\Lambda_{\bf k}-\hbar{\bf
  k}\cdot {\bf v}.\nonumber
\end{eqnarray}

Thus, we have gathered all the ingredients for conducting the transport
calculations via the diagrammatics of the \emph{replica trick}.  The above
model is quite general and allows for calculating observables in a
broad range of scenarios by including appropriate orders in the
perturbation expansion. For example, if the time scale of interest is
larger than the Heisenberg time (time scale defined by the trapping
potential), one needs to include appropriate particle-hole diagrams to
describe the diffusive dynamics. On the other hand, if the dynamics
occurs on the trap scale, then depending on the disorder strength,
appropriate single particle diagrams may be sufficient. The only point
one needs to remember is that the bare propagators of the replica
theory are those for the quasiparticles representing atoms dressed by
interactions with the corresponding Green's function given by~\cite{ripka,mahan}:
 \begin{eqnarray}
\mathscr{G}({\bf k},i\omega_n)[{\bf r}]=\frac{1}{\Lambda_{{\bf k}}^{{\rm lab}}[{\bf r}]-i\omega_n}+\frac{1}{\Lambda_{{\bf k}}^{{\rm lab}}[{\bf r}]+i\omega_n},\nonumber
\end{eqnarray}
where $\omega_n$ is the bosonic Matsubara frequency.

\section{Experiment} As elaborated in the introduction, our
model is motivated by recent experiments performed on cold trapped
atomic condensates. Hence, we illustrate the applicability of
the model by considering the actual experimental situation of
Refs.~\cite{randy,randy2}: $^7{\rm Li}$ atoms trapped in an axial
harmonic trap with $\omega_z=2\pi\times 5.5\,\, {\rm Hz}$, aspect
ratio $\alpha= \omega_{\perp}/\omega_z \approx 46$. The $s$-wave
scattering length is tuned via a Feshbach resonance to about
$a=25\,a_{{\rm B}}$, resulting in an axial condensate with the size
represented by the Thomas-Fermi (TF) radius, which in units of the
axial oscillator length $\ell_z$, is given by $R\approx 13.6
\,\ell_z$.  These parameters conform with the experimental data for
damping of the BEC dipole mode used in the plot of
Fig.~\ref{oscillations}. The disorder potential is produced using an
optical speckle with gaussian correlation:
\begin{eqnarray}
\langle
U_{{\rm d}}({\bf x})U_{{\rm d}}({\bf
  x}')\rangle_{{\rm dis}}&=&V_{{\rm d}}^2\exp\left[-\frac{2({\bf x}-{\bf
  x}')^2}{\sigma^2}\right].\label{disorderfunction}
\end{eqnarray}
Further, while no data on temperature is available, from the
observations presented, it is safe to assume that the experiment is
conducted at temperatures well below $T_{{\rm BEC}}$ and thus a $T=0$
theoretical prescription should suffice.
%, implying $n_{{\bf k}}=0$ for ${\bf k\ne 0}$.

The TF density of the stationary condensate in cylindrical
coordinates follows from Eqs.~\eqref{lda1} and \eqref{lda2}
\begin{eqnarray}
\rho(r_{\perp}
,z)&=&\frac{m\omega_z^2(R^2-\alpha^2r_{\perp}^2-z^2)}{2\lambda},\nonumber
\end{eqnarray}
with the normalization fixed by the TF axial size $R$ via
\begin{eqnarray}
N=4\pi\int_0^{R/\alpha} \int_{0}^{R[r_{\perp}]}\rho(r_{\perp},z){\rm d}z\,\,r_{\perp}{\rm d}r_{\perp},\label{normalization}
\end{eqnarray}
with the different distances indicated in
Fig.~\ref{distances}. For convenience, we define the dimensionless
ratio $\xi[r_{\perp}]=v/c[r_{\perp}]$, where $c[r_{\perp}]$ is the
speed of sound at the coordinate $\{r_{\perp},z=0\}$. We also use the
notation, $c\equiv c[0]$ and $\xi\equiv\xi[0]$. We point out that even
though the aspect ratio is large, the chemical potential $\mu\sim{\cal
  O}(\hbar\omega_{\perp})$, hence a full 3D treatment is
required. Finally, we observe that the oscillation data shown in
Fig.~\ref{oscillations} indicates
dynamics over time scales much shorter than the diffusion time
dictated by the disorder. This greatly reduces the manipulations and it
suffices to calculate only the single particle diagrams.
%giving the
%dressed propogator of the non-interacting quasiparticles.
The imaginary part of the selfenergy, obtained from the dressed
propagator of the non-interacting quasiparticles, immediately provides
the transition rate $W({\bf k}\rightarrow {\bf k}',{\bf r})$ at the
spatial point ${\bf r}$ (for details refer to the Appendix). However, we alert the reader that one
typically never measures the single particle propagator in an
experiment and we still need to connect to the damping of the
classical dipole mode. We achieve this by recalling that the energy
per condensate particle is nothing but the chemical potential
$\mu({\bf r})$. Thus, the total energy transfer rate out of the ${\bf
  k=0}$ condensate mode is precisely
\begin{eqnarray}
\Gamma=\int_{\Omega} {\rm d}{\bf r}\int_{k>\kappa}{\rm d}{\bf k}\,\,\mu({\bf r})W(0\rightarrow {\bf k},{\bf r}),\label{energylossrate}
\end{eqnarray}
where $\kappa$ is some infrared cutoff imposed by the trapping
potential and $\Omega$ is the condensate volume. The above integral
can be approximated and has a nice analytic form if excitations are
allowed only along the axial (z) direction. This constraint can be
understood to arise from the high aspect ratio $\alpha$ of the
trapping potential. Even at large $k$, where this approximation may
fail, we expect the density of states to be diminished enough to
accrue any appreciable error. With this we arrive at the final result
in terms of dimensionless quantities represented by
$(\bar{\,\,\,\,})$(physical quantities scaled in units of the axial
harmonic trap: lengths in units of $\ell_z=\sqrt{\hbar/m\omega_z}$ and
energies in $\hbar\omega_z$)
\begin{eqnarray}
\bar{\Gamma}[\xi]=\frac{\bar{\gamma}^2\bar{R}^4}{4\sqrt{2}\pi\bar{a}}\int_0^1 t^{2}\,\,
{\cal I}[\xi/\sqrt{t},\xi_{{\rm min}}/\sqrt{t}]\,\,{\rm d}t, \label{main1}
\end{eqnarray}
where the upper (lower) integration limit corresponds to the center
(radial edge) of the condensate, $\xi_{{\rm min}}$ is related to
$\kappa$, $\xi_{{\rm min}}=\hbar\kappa/2mc$ and the function ${\cal
  I}$ is defined by Eqs.~\eqref{boundaryfunction} and \eqref{function} of Appendix.

\section{Results and Discussion} 
\begin{figure}[t]
 \includegraphics[scale=.43]{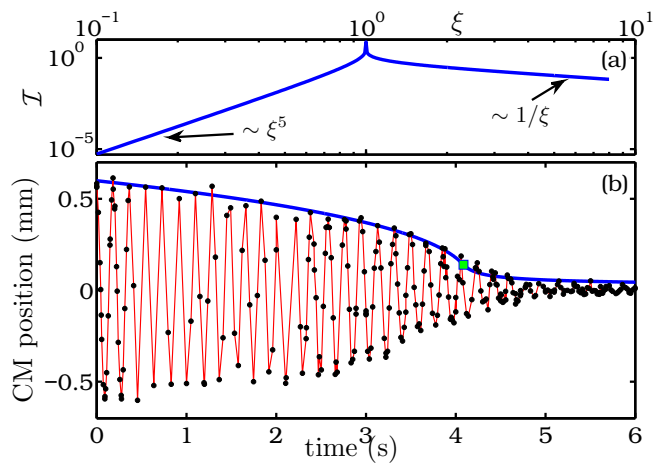}
  \caption{(Color online) (a) Damping function along the long axis of the condensate
    plotted as a function of $\xi$.  (b) Thick solid curve:
    theoretical prediction for the peak center-of-mass displacement of the
    condensate as a function of time for $\bar{\gamma}=0.5$, filled
    circles connected with thin line are experimental data points for
    the center-of-mass position as the condensate oscillates in the trap
    \cite{randy2}. Square marker indicates the point where $\xi_{{\rm peak}}=1$.}
\label{oscillations}
\end{figure}

To begin with, we first look at the
plot in Fig.~\ref{oscillations}(a) that shows the velocity dependence
of the function ${\cal I}$ for points along the long axis of the
condensate. The sharp discontinuity near the speed of sound $c$ is a
clear indication of the origin of distinct regimes of damping as the
BEC flows through the disordered medium. Essentially, for velocities
significantly smaller than the speed of sound $c$, excitations are
only allowed in a small shell near the surface of the condensate. This
region tends to grow inwards, eventually allowing excitations
everywhere in the condensate, as the center-of-mass velocity
reaches $c$. However, increasing the velocity beyond this critical
value, the excited volume remains fixed while the Bogoliubov density
of states keeps falling, resulting in net decay of the
damping rate with velocity.

The above expression for $\bar{\Gamma}$ represents the first of the two main
results of this article. It conveniently leads us to a dynamical
equation for the condensate motion, if we picture the BEC oscillating
in the random potential as a rigid body whose hydrodynamic
oscillations are damped as the kinetic energy is transferred to the
internal excitations. Such damped oscillatory motion of the BEC center-of-mass
coordinate can easily be described by an equation for the peak
velocity of the condensate motion in terms of $\xi_{{\rm peak}}$
\begin{eqnarray}
  \frac{\partial\xi_{{\rm peak}}}{\partial t}= -\frac{2}{N\bar{R}^2 \xi_{{\rm peak}}}\tilde{\Gamma}[\xi_{{\rm peak}}]=-\beta \xi_{{\rm peak}},
  \label{main2}
\end{eqnarray}
where the energy dissipation rate is averaged over one oscillation
period, represented by $\tilde{\Gamma}=(1/T)\int_0^T\Gamma {\rm d}t$.
The above equation for $\xi_{{\rm peak}}$ represents the second main
result of this article. These results immediately provide important
asymptotic exponents governing the dependence of the velocity-damping rate
$\beta$ on the scattering length. For this, all that is required is to
remember that the total number of atoms $N$ is fixed, automatically
implying $R\sim a^{1/5}$ from the normalization condition Eq.~
\eqref{normalization}; thus, $c\sim a^{1/5}$ follows as well. Now, from the
functional form given by Eq.~\eqref{function},
\begin{eqnarray}
\beta \sim \left\{
\begin{array}{l}
a^{-6/5} \,\,\,\,\forall\,\,\xi\ll 1 \Rightarrow {\cal I}\sim 1/a\\
a^0 \,\,\,\,\,\,\,\,\,\,\,\,\,\forall\,\,\xi\gg 1
\end{array}
\right.
\end{eqnarray}
The latter limit is easily confirmed from the plot,
Fig.~10, in Ref. \cite{randy2}. While, data for the former limit
is not present in the same figure, available data points for $\xi <1$
show a clear trend towards the predicted limiting behavior.  Instead,
we indirectly verify this limit by extracting the damping rate as a
function of $\xi$ from the data shown in Fig.~\ref{oscillations}(b)
which can then be written as a function of scattering length, $a$, by
fixing $v$ in Eq.~\eqref{main1}. We again find very good agreement,
confirming the exponent corresponding to the former limit as well.

The only parameter that remains to be estimated is the
$\delta$-correlated disorder strength $\gamma$. We remind the reader
that, in the actual experiment, the disorder potential, Eq.~
\eqref{disorderfunction}, has a finite correlation length $\sigma$.
Such spatial correlations are indeed outside the scope of the
current LDA based model. Therefore, $\gamma$ in our model maybe
considered as a free parameter representing the renormalized value of
the disorder strength.
% shown in Ref.~\cite{randy2}.
Our result, the solution of Eq.~\eqref{main2} plotted in Fig.~\ref{oscillations}(b), is in excellent
quantitative agreement with the experimental data, indicating the
validity of a delta-correlated ansatz for the speckle potential. The
proposed model is also able to capture all the damping time scales via
only coupling to the Bogoliubov excitations. Thus, we identify the
most dominant source of dissipation, sufficient to describe transport
in such ultra-cold trapped-atom experiments.

Finally, we shall comment on the strength of the disorder. For this,
let us use the value of $\bar{\gamma}$ estimated in
Fig.~\ref{oscillations}(b) to find the magnitude of disorder-induced
condensate depletion given via the method of Ref.~\cite{huang}
\begin{eqnarray}
\rho_{{\rm dep}}(0)&=&\frac{m^2\gamma^2}{8\pi^{3/2}\hbar^4}\sqrt{\frac{\rho(0)}{a}}. \nonumber
\end{eqnarray}
We find $\rho_{{\rm dep}}(0)/\rho(0)=.002$, implying
that the condensate density remains almost unchanged after inclusion
of speckle potential and thus justifying the assumption of weak disorder,
permitting perturbative treatment up to second order in disorder strength.
Furthermore, it
also points to an important aspect about the disorder potential used
in the experiment~\cite{randy2}. There, even though the disorder
strength seems to be large, given by $V_{{\rm d}}\approx 0.6\,\mu$,
our theoretical model points to a critical interplay between disorder
correlations and interactions, resulting in a dressed disorder with a
small renormalized strength $\gamma$.

\section{Conclusion}
To summarize, we have given a theory for describing the transport
properties of \emph{finite-size} atomic condensates flowing through
disorder potentials. We emphasize the intricate role played by the
inhomogeneous character of the condensate, resulting in a completely
different form of the damping function, ${\cal I}$, above and below
the critical velocity, $c$. The validity of our model is established
via comparison with the experimental observations of damped
hydrodynamic dipole oscillations of the condensate \cite{randy2}. The
predicted power-law dependance of the damping on the interaction parameter is
in excellent agreement with experimental observations. Our
results, while based on a white noise disorder model, show extremely
good agreement with experimental data, thereby, indicating the
insignificant role of finite correlations in determining the
dissipation time scales.  While a full theory was not necessary for
the aspects considered here, an improved quantitative picture can be
realized by including finite disorder-correlations, for example, via
a coherent potential approximation \cite{cpadig}. Such procedure leads
to an effective, renormalized disorder vertex, whose value can then be
compared with the $\gamma$ obtained here. However, these calculations
are beyond the current model and would be considered in a future work.
Furthermore, although the effects of Anderson localization turn out not to be relevant
for these experiments\cite{randy,randy2}, we predict they could be if the center-of-mass
motion is very slow and therefore the physics is dominated by diffusion.
These effects could be incorporated into our calculations by including appropriate
diagrams corresponding to particle-hole processes. Finally, the analysis presented here can be easily translated in the language of
friction indicating the manifestation of non-linear friction on mesoscopic quantum objects.

\begin{acknowledgments}
We are extremely
grateful to Randy Hulet, Scott Pollack, and Dan Dries for invaluable
discussions relating to their experimental findings on the dissipative
transport. We acknowledge the financial support from the W. M. Keck
Foundation and NSF.  C.J.B. and P.K. also acknowledge the financial
support from ARO Award W911NF-07-1-0464 with the funds from the DARPA
OLE Program.
\end{acknowledgments}
\appendix
\section{Replica Method} The difficulty of disorder averaging is
eliminated using the standard \emph{replica trick} of considering $R$
replicas of the same field with the understanding that the
$R\rightarrow 0$ limit is taken at the end. However, this procedure
comes with a cost, it leads to an effective attractive interaction
between atoms from different replicas. To illustrate this, we simply
write down the full average of some operator $\hat{{\cal O}}$, as
\begin{eqnarray}
\langle\hat{{\cal
    O}}\rangle\hspace{-.1cm} &=&\lim_{R\rightarrow
  0}\left[\frac{1}{R}\frac{\partial \langle {\cal Z}^R[\eta]\rangle_{{\rm dis}}}{\partial \eta}\right],
  \label{replica_limit}
\end{eqnarray}
where $\eta$ is some source field, and ${\cal Z}^R[\eta]$ is the
partition function of the replicated boson field
$\{(\psi^{a\dagger},\psi^a),a=1,..,R\}$. Thus, essentially it involves
an averaging of the partition function over the distribution,
$P[U_{{\rm d}}]$, that can be carried out quite easily using Gaussian
integration resulting in an effective action
\begin{eqnarray}
S_{{\rm eff}}&=&\sum_{a=1}^RS_{{\rm clean}}[\psi^a,\psi^{a\dagger}]+\sum_{a,b=1}^RS_{{\rm replica}}[\psi^a,\psi^b,\psi^{a\dagger},\psi^{b\dagger}],\nonumber
\end{eqnarray}
with
\begin{eqnarray}
S_{{\rm replica}}&=&-\frac{\gamma^2}{2}\sum_{m,n}\int{\rm d}{\bf x}~\psi_m^{a\dagger}({\bf x})\psi_m^{a}({\bf x})\psi_n^{b\dagger}({\bf x})\psi_n^{b}({\bf x}),\nonumber
\end{eqnarray}
where $S_{{\rm clean}}$ is the action in the absence of disorder, and index $m$
and $n$ are introduced to indicate Matsubara frequencies $\omega_m$
and $\omega_n$, respectively. Thus, the additional replica-induced
action represents interactions between atoms of different replica
index with the disorder vertex represented by the Feynman diagram of Fig.~\ref{feyn}(a). Also, it does not involve energy exchange between replicas
since the disorder potential is intrinsically time independent.

\section{Damping Function}
\begin{figure}[ht]
  \includegraphics[scale=.48]{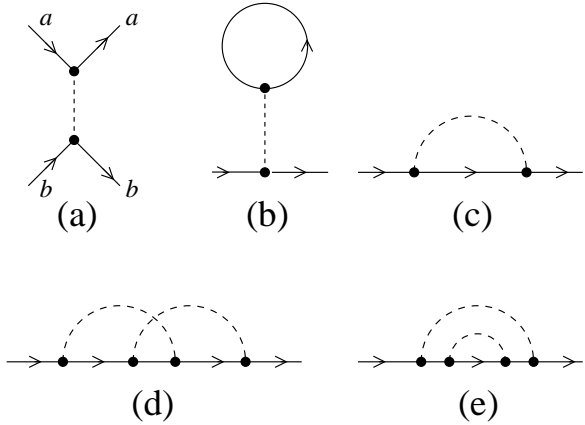}
  \caption{Diagram contributions to the self energy: (a) corresponds to the disorder vertex, $m$ and $n$ are replica index. (b) and (c) are second order,
  (c) and (d) are fourth order contributions to the selfenergy. The external lines are not to be included for selfenergy calculations.
  The solid lines represent the quasiparticle propagator.}
\label{feyn}
\end{figure}

The relevant diagrams are the single particle ones as shown in Fig.~\ref{feyn}(b)-(e).
A technical point of interest is that the contribution of diagrams
containing loops of the type shown in Fig.~\ref{feyn}(b) tend to zero due to the limit $R \rightarrow 0$ in Eq.~\eqref{replica_limit}. In essence, the replica trick
eliminates all diagrams that are disconnected before the
disorder averaging, thus vastly reducing the computation.

For the particular experimental situation considered here, it suffices to include just the leading order diagrams. Further,
the $T=0$ nature of the experiment makes the use of \emph{replica trick} redundant and the result coincides with the Fermi's golden rule.
However, we remind the reader that the model itself is more general and maybe applied to other situations as mentioned
in the text. All throughout, our goal has been to provide a theoretical model to describe transport in trapped, disordered, ultra-cold atomic experiments in general.
Thus, the formal expression for the transition rate is given simply by the imaginary part of selfenergy (diagram of Fig.~\ref{feyn}(c)).
\begin{eqnarray}
  W(0\rightarrow {\bf k},{\bf r})=\frac{2\pi\gamma^2}{\hbar}\rho({\bf r}) (u_{\bf k}[{\bf r}]-v_{\bf k}[{\bf r}])^2 \delta(\Lambda_{\bf k}[{\bf r}]-\hbar {\bf k}\cdot {\bf v}),\nonumber
\end{eqnarray}
where $u_{{\bf k}}$ and $v_{{\bf k}}$ are the amplitudes that define
Bogoliubov quasiparticles and are related to the energy $\Lambda_{{\bf
    k}}=\sqrt{e_k^2-\lambda^2\rho^2}$ by $u_{{\bf k}}^2=(e_k/\Lambda_k+1)/2$ and $v_{{\bf
    k}}^2=(e_k/\Lambda_k-1)/2$ with $e_k=\hbar^2k^2/2m+\lambda\rho$. Therefore, the full expression for the
loss rate $\Gamma$ can be written simply by substituting in
Eq.~\eqref{energylossrate} of text. Performing the $z$ and the $k$ integration
we arrive at
\begin{eqnarray}
\Gamma&=&\frac{2\pi\gamma^2m^2\omega_z^3}{\pi\sqrt{2}\hbar^2\lambda R_{\perp}^2}\int_0^{R/\alpha} r_{\perp}{\rm d}r_{\perp}R[r_{\perp}]^4{\cal I}[\xi[r_{\perp}],\xi_{{\rm min}}[r_{\perp}]],\nonumber
\end{eqnarray}
where $R[r_{\perp}]$, indicated in Fig.~2, is written as
$R[r_{\perp}]=\sqrt{R^2-\alpha^2r_{\perp}^2}$ and the function ${\cal
  I}$ is defined in terms of the boundary function ${\cal F}$
\begin{equation}
{\cal I}[p,q]\hspace{-.1cm}=\hspace{-.1cm}\left\{\hspace{-.1cm}
\begin{array}{lr}
{\cal F}(1,p)-{\cal F}(0,p),&\hspace{-.1cm}\forall p \ge \sqrt{1+q^2}\\
{\cal F}(1,p)-{\cal F}(\sqrt{1-p^2+q^2},p),&\hspace{-.1cm}\forall p < \sqrt{1+q^2}
\end{array},
\right.\label{boundaryfunction}
\end{equation}
with
\begin{eqnarray}
  {\cal F}(y,x)&=&\frac{1}{8}\left[ y(-5+2y^2-3x^2)\sqrt{-1+x^2+y^2}+\right.\nonumber\\
  &&\hspace{-1.1cm}\left.(3+2x^2+3x^4)\log\left(2y+2\sqrt{-1+y^2+x^2}\right)\right].
  \label{function}
\end{eqnarray}
Now, with straightforward manipulation, one can immediately show that
Eq.~\eqref{main1} of text follows by simply expressing all quantities in
dimensionless units.

\end{document}